\documentclass[preprint,showpacs]{revtex4}
\usepackage{amssymb}
\usepackage{graphicx}
\begin{document}
\title{Surface and Bulk Properties of Deposits Grown with a Bidisperse Ballistic
Dposition Model}
\author{F. A. Silveira${}^{1,}$\footnote{Email address: fsilveira@if.uff.br (corresponding author)} and
F. D. A. Aar\~ao Reis${}^{2,}$\footnote{Email address: reis@if.uff.br}}
\affiliation{Instituto de F\'\i sica, Universidade Federal
Fluminense, Avenida Litor\^anea s/n, 24210-340 Niter\'oi RJ, Brazil}

\date{\today}

\begin{abstract}
We study roughness scaling of the outer surface and the internal porous
structure of deposits generated with the three-dimensional bidisperse ballistic
deposition (BBD), in
which particles of two sizes are randomly deposited. Systematic extrapolation
of roughness and dynamical exponents and the comparison of roughness
distributions indicate that the top surface has Kardar-Parisi-Zhang scaling for
any ratio $F$ of the flux between large and small particles. A scaling theory
predicts the characteristic time of the crossover from random to correlated
growth in BBD and provides relations between the amplitudes of roughness
scaling and $F$ in the KPZ regime. The porosity of the deposits monotonically
increases with $F$ and scales as $F^{1/2}$ for small $F$, which is also
explained by the scaling approach and illustrates the possibility of connecting
surface growth rules and bulk properties. The suppression of relaxation
mechanisms in BBD enhances the connectivity of the deposits when compared to
other ballisticlike models, so that they percolate down to $F\approx 0.05$.
The fractal dimension of the internal surface of the percolating deposits is
$D_F\approx 2.9$, which is very close to the values in other ballisticlike
models and suggests universality among these systems. 
\end{abstract}

\maketitle

\section{Introduction}
\label{intro}

The large number of applications of porous materials and the need
to understand and control their properties motivates the proposal of
models to represent their geometry and the study of its effects on physical
processes taking place in the pore system \cite{hilfer,sahimi}. In some models,
the formation of the porous structure is a consequence of a certain growth
dynamics which represents its main mechanisms. One example is
ballistic deposition (BD), which was originally proposed by Vold to describe
sedimentary rock formation \cite{vold}, and whose growth rules are illustrated
in Fig. 1(a). Extensions of this model were already used to produce porous
deposits where fluid adsorption and diffusion were studied in connection to the
bulk geometry (see, e.g., Ref. \protect\cite{kikkinides}). In parallel, surface
roughness of ballistic deposits have also attracted much attention
\cite{ball,baiod,raissa,balfdaar,katzavbd,perez,haselwandter,extremebd,extremebd1},
particularly for the connection with the Kardar-Parisi-Zhang (KPZ) theory
\cite{kpz}.

A related model, called bidisperse ballistic deposition (BBD), was recently
proposed for sedimentary rock formation. Its growth rules are illustrated in
Fig. 1(b). In the three-dimensional version, particles of two different sizes may
incide vertically towards the surface: single site particles (size $1\times
1\times 1$ in lattice units) with probability $1-F$ and double site particles
(size $2\times 1\times 1$) with probability $F$. Any incident particle
permanently sticks at the first position where it encounters a previously
deposited particle below it. This bimodal distribution of particle size is
believed to be realistic because real sand grains are reported to be
ellipsoidal with the long axis approximately twice the shortest one
\cite{bbd1}. Indeed, there is a large number of simple growth models with two
species of particles or two particle orientations which can reproduce real
systems features \cite{kikkinides,guyon,tetris,poison1,heapmodel,trojan}, which
justifies the interest in models such as BBD.

Due to its simple stochastic rules, BBD may be useful to connect the growth
mechanisms, which depend only on the outer surface properties, and the internal
morphology of porous deposits. However, this important question was not
addressed in previous work on BBD. In Ref. \protect\cite{bbd1}, some properties
of the porous deposits were numerically studied, such as porosity and
permeability, but detailed information on their connectivity was not provided.
In Ref. \cite{bbd2}, where only surface roughness scaling was analyzed, it was
suggested that the dynamical exponent abruptly changes from the KPZ value for
$F>0.2$ to another value for smaller $F$ (this is typical of a transition
between different dynamic growth phases \cite{poison1}). 

This scenario motivates the present work, in which a systematic analysis of
surface and bulk properties of the deposits generated by BBD in $2+1$
dimensions will be performed. Our first step is to show that deviations of
scaling exponents from the KPZ values are related to a crossover from random to
correlated growth, whose characteristic times are very
large for small $F$. This conclusion follows from a scaling approach similar to
Refs. \protect\cite{rdcor,rdcoralbano,lam} and  will also be confirmed by
simulation results. The comparison of scaled roughness distributions of BBD
and other KPZ models \cite{distrib} will provide additional support to the claim
that BBD is in the KPZ class for any value of $F$.

However, surface roughness scaling contains limited information on the structure
of the deposit generated by BBD because it only
takes into account the correlations of the heights of the highest particles of
each column of the deposit, i.e., of the outer surface. We will show that the
pore system below this surface has a high connectivity (i.e., the system of
pores percolates) in an exceptionally large range of values of $F$, from $F=1$
down to $F\approx 0.05$, so that the internal surface of the deposit has a
fractal dimension as large as $D_F\approx 2.9$. This value is very close to
$D_F$ for the original BD model and for an extension of BD including relaxation
\cite{yu1}, which suggests universality of the surface fractal dimension among
a large class of ballisticlike deposits.

In addition, we will illustrate the possibility to connect the internal
properties of the deposit with growth rules (the particle size distribution and
the parameter $F$) by extending our scaling arguments to show that the porosity
of the deposits scales as $F^{1/2}$ for small $F$. Only a small number of
recent works was devoted to the simultaneous study of bulk and surface
properties in growth models \cite{strucbelow,yu1,yu2}, but such studies are
certainly very important for a realistic modelling of growth of porous media. 

The rest of this paper is organized as follows. In Sec. II we present estimates
of roughness and dynamical exponents for the BBD and compare numerically
calculated roughness distributions in the steady states. In Sec. III we present
a scaling theory that explains the crossover from random to KPZ growth in the
BBD model, with the relative flux $F$ being the control parameter of this
crossover. In Sec. IV we analyze the porosity of the deposits. In Sec. V we
study the connectivity of the deposits and the scaling properties of the
internal surface. In Sec. VI we summarize our results and discuss possible
applications and extensions of the present work.

\section{Numerical study of roughness scaling}

In surface growth models, the roughness $W$ is usually defined as the
rms fluctuation of the height $h$ around its average position $\overline{h}$:
\begin{equation}
W(L,t)\equiv {\left[ { \left<  {\left( h - \overline{h}\right) }^2  \right> }
\right] }^{1/2} ,
\label{defw}
\end{equation}
where the overbars indicate spatial averages and the angular brackets indicate
configurational averages.
In models with pore formation, such as BD and BBD, many particles at each
substrate column may be in contact with the external media, thus the interface
of the solid phase is multivalued. However, in the definition of the roughness
$W$, the height $h$ of each column is chosen as the height of the highest
particle in that column. Thus, $W$ characterizes the outer surface of the
deposit, which may be accessible through surface imaging methods and whose
scaling properties may be useful for many applications \cite{barabasi}.

The above defined roughness certainly do not represent the properties of the
porous media, which has a much more complex internal surface, to be studied in
Secs. IV and V. In systems with multivalued interfaces and anomalous scaling of
$W$  (which is not the case of BBD), it may even be difficult to characterize
the surface properly by calculating this quantity -- see, e.g., Ref.
\protect\cite{bru}. 

In a random, uncorrelated growth, the roughness increases as
\begin{equation}
W_\textrm{\footnotesize random}\sim t^{1/2} .
\label{wrandom}
\end{equation}
This is the case of the BBD with $F=0$, in which the deposition in a given
column is independent of the neighboring ones. However, in correlated growth
processes, the roughness
is expected to obey the Family-Vicsek scaling relation \cite{fv}
\begin{equation}
W(L,t) \approx A L^{\alpha} f\left( \frac{t}{t_\times}\right) ,
\label{fv}
\end{equation}
where $L$ is the system size, $\alpha$ is the roughness exponent, $A$ is a
model dependent constant, $f$ is a scaling function such that $f\sim 1$ in the
regime of roughness saturation
($t\to\infty$), and $t_\times$ is the characteristic time of crossover to
saturation. $t_\times$ scales with the system size as 
\begin{equation}
t_\times \approx BL^z , 
\label{scalingtau}
\end{equation}
where $z$ is the dynamic exponent and $B$ is another model dependent
amplitude. For $t\ll t_\times$ (after a possible crossover to correlated
growth), the roughness scales as
\begin{equation}
W\approx Ct^\beta , 
\label{scalingwgr}
\end{equation}
where $C$ is also model dependent and $\beta=\alpha/z$ is the growth exponent.
In this growth regime, $f(x)\sim x^\beta$ in Eq. (\ref{fv}). In most
limited-mobility growth models (i.e., models without collective diffusion), the
time unit is taken as the time necessary to deposit the mass of one monolayer.

The rms fluctuation of the squared roughness in the steady state is
\begin{equation}
\sigma\equiv \sqrt{ \left< {w_2}^2 \right> - {\left< w_2\right>}^2 } ,
\label{defsigma}
\end{equation}
where $w_2\equiv \overline{h^2} - {\overline{h}}^2$ indicates the roughness of a
given steady state configuration. $\sigma$ scales with the same exponents of $W$ multiplied
by $2$, but with much weaker finite-size corrections, as previously shown for
some growth models in Ref. \protect\cite{distrib}. Thus, here we will use
$\sigma$ instead of $W$ for estimating roughness exponents. Effective roughness exponents are 
defined as
\begin{equation}
\alpha\left( L\right) \equiv {1\over 2} {
\ln{ \left[\sigma\left( L\right)
/ \sigma\left( L/2\right)\right] }\over \ln{2} } .
\label{defalphaef}
\end{equation}

We simulated the BBD model for
several values of $F$, ranging from $F=0.06$ to $F=0.40$, in two-dimensional
substrates of linear sizes $L$ ranging from $L=16$ to $L=512$, up to the steady
states. Nearly ${10}^4$ configurations were simulated for each $F$ and $L$.
In Fig. 2(a) we show the effective exponents $\alpha\left( L\right)$ as a
function of $1/L$ for $F=0.08$. Extrapolation to $L\to\infty$ gives $\alpha=
0.395$. The variable in the abscissa was shown to provide a good linear fit of
the large $L$ data among other integer and half-integer powers of $L$.
A more refined extrapolation procedure, similar to  Ref. \protect\cite{kpz2d},
gives nearly the same value of $\alpha$ as $L\to\infty$.

The value $F=0.08$ in Fig. 2(a) is in the region where KPZ scaling was
not observed in previous work \cite{bbd2}. However, the extrapolated $\alpha$ is
in excellent agreement with the best known estimates for the KPZ class in
$d=2+1$ dimensions, which are in the range $\left[ 0.375, 0.396\right]$
\cite{kpz2d}. Our estimates of $\alpha$ for all values of $F$, from $0.06$ to
$0.40$, are also consistent with the KPZ range.

We also calculated characteristic times $\tau(L)$ which are proportional to the
saturation times $t_\times$ (Eqs. \ref{fv} and \ref{scalingtau}), using
the method introduced in Ref. \protect\cite{tau}. Effective exponents $z$ are
defined as 
\begin{equation}
z\left( L\right) \equiv {
\ln{ \left[\tau \left( L\right)
/ \tau\left( L/2\right)\right] }\over \ln{2} } .
\label{defzef}
\end{equation}
In Fig. 2(b) we plot $z(L)$ as a function of $1/L$ for $F=0.1$, which gives
$z=1.66\pm 0.05$ asymptotically. This value of $F$
is also in the region where Ref. \protect\cite{bbd2} did not find KPZ scaling.
However, the above estimate also has an intersection with the best previous
estimate of $z$ for the KPZ class, in the range
$\left[ 1.605,1.64\right]$ \cite{kpz2d}. For the other values of $F$, the
estimates of $z$ also intercept this KPZ range.

Additional support to the claim that BBD has KPZ scaling follows from comparison
of distributions of the squared width $w_2\equiv \overline{h^2} -
{\overline{h}}^2$ in the steady state. Letting $P\left( w_2\right)$ be the
probability density of the roughness of a given configuration to lie in the
range $\left[ w_2, w_2+dw_2\right]$, it is expected that this density satisfies
the scaling relation $P\left( w_2\right) = {1\over \sigma}
\Psi\left( {{w_2-\left< w_2\right>}\over\sigma}\right)$, with $\sigma$ defined
above and $\Psi$ being a universal function (see, e.g., Ref.
\protect\cite{antal}). In Fig. 3 we plot the steady state
scaled distributions for the BBD model with $F=0.06$ and $F=0.1$ and for the
restricted solid-on-solid (RSOS) model, which is one of the best
representatives of the KPZ class. The good collapse of the curves for the
different models indicates that BBD is in the KPZ class in $2+1$ dimensions for
all values of $F$.
Quantitatively, the collapse of the curves is confirmed by the values
of the skewness $S$ and of the kurtosis $Q$: for BBD with $F=0.06$ we have
$S=1.70\pm 0.02$
and $Q=5.50\pm 0.20$, for BBD with $F=0.1$ we have $S=1.69\pm 0.02$ and
$Q=5.42\pm 0.20$, 
and for the RSOS model we have $S=1.71\pm 0.02$ and $Q=5.4\pm 0.1$
\cite{distrib}.

\section{Scaling theory for the crossover of surface roughness}

In order to provide an explanation for the deviations in the roughness
scaling of BBD observed in previous works, now we analyze its scaling
properties along the same lines of other competitive models
\cite{rdcor,rdcoralbano,lam}. This approach will also be useful to explain the
scaling of the porosity for small $F$ in Sec. IV.

In BBD with high fluxes of large particles ($F\lesssim 1$), the ballisticlike
nature of the problem is clear and the KPZ scaling was shown in early simulation
work \cite{bbd2}. On the other hand, with small $F$, most deposition attempts
(of small
particles) lead to uncorrelated growth. However, as shown in Fig. 1, a single
deposition of a large particle (whose long axis is oriented in one of the
surface directions) always leads to the same final height in two neighboring
columns. The aggregation of this particle  immediately introduces correlations
among those columns. The typical time interval for
deposition of this particle in a given column is $1/F$, which is large when $F$
is small, while that typical time is of order unity for $F\approx 1$.
Consequently, we expect that
the same features of the model with $F\approx 1$ will be present in the model
with small $F$, but with all characteristic times rescaled by a factor $1/F$.

The constant $B$ in Eq. (\ref{scalingtau}) is of order
$1$ for the model with $F\approx 1$, as well as in the original BD model (see, e.g.,
 Ref. \protect\cite{tau}). Thus, in BBD with small $F$, we expect that
\begin{equation}
B\sim 1/F .
\label{BBBD}
\end{equation}

Now consider a narrow system, i.e., with lattice size $L$ of order unity. The
columns inside this system randomly grow until a time of order
$1/F$. During this time
interval, the roughness increases as in random deposition (Eq. \ref{wrandom}).
Thus, at $t\sim 1/F$, the roughness
is of order ${\left( 1/F\right)}^{1/2}$. When a small number of correlated
depositions occurs, the whole (small) system will be correlated and the
roughness will saturate at a value of this order of magnitude. This means that
the amplitude in Eq. (\ref{fv}) must scale as
\begin{equation}
A\sim 1/F^{1/2}
\label{ABBD}
\end{equation}
in BBD with small $F$. Combined with this result,
FV scaling and Eq. (\ref{scalingwgr}) immediately lead to $C\sim
1/F^{1/2-\beta}$, with the KPZ exponent $\beta$.

The above discussion shows that deviations from KPZ scaling in BBD with small
$F$ are just finite-time or finite-size effects due to long crossovers from
random to correlated growth. In order to test this scaling approach, we
estimated the amplitudes $A$ and $B$ in the limit $L\to\infty$ by extrapolating
the ratios $W_\textrm{\footnotesize sat}(L)/L^\alpha$ and $\tau(L)/L^z$, respectively. The values
$\alpha= 0.385$ and $z=1.615$ of the KPZ class \cite{kpz2d} were used in the
calculation of those quantities. Following the notation of Ref.
\protect\cite{horowicz}, we assume that
\begin{equation}
A\sim F^{-\delta}
\label{defdelta}
\end{equation}
and 
\begin{equation}
B\sim F^{-y} .
\label{defy}
\end{equation}
Effective exponents for $\delta$ and $y$ were calculated as
\begin{equation}
\delta(F) \equiv -{
\ln{ \left[A \left( F''\right)
/ A\left( F'\right)\right] }\over \ln{(F''/F')} },
\label{deltaeff}
\end{equation}
\begin{equation}
y(F) \equiv -{
\ln{ \left[B \left( F''\right)
/ B\left( F'\right)\right] }\over \ln{(F''/F')} },
\label{yeff}
\end{equation}
with $F=\sqrt{F''F'}$ and successive values $F'$, $F''$.
In Figs. 4(a) and 4(b) we show $\delta(F)$ and $y(F)$
as functions of $F^{1/2}$ and $F$, respectively, and linear extrapolations of
the data for small $F$. Again, the variables in the abscissas were chosen to
provide good linear fits of the low $F$ data. The extrapolations to $F\to 0$
give exponents consistent with the
values $\delta=1/2$ and $y=1$ predicted in Eqs. (\ref{ABBD}) and (\ref{BBBD}),
respectively, thus confirming the validity of our scaling approach. 

\section{Porosity of the deposits}

Now we begin the analysis of the internal structure of the deposits.
The simplest quantity to characterize a porous structure is the porosity $P$,
which is the fraction of empty lattice sites inside the deposit. In Fig. 5 we
show a log-log plot of $P$ as a function of $F$. Results for different substrate
sizes confirm the absence of significant finite-size effects. As $F$ increases
towards $F=1$, $P$ tends to saturate, as expected. However, for small $P$, a
fit of our data gives
\begin{equation}
P\sim F^a ,
\label{scalingP}
\end{equation}
with a numerical estimate $a=0.503\pm 0.003$ obtained from the data for $L=256$.
This result contrasts to the exponent $a=0.63$ obtained in Ref.
\protect\cite{bbd1} in $2+1$ dimensions (in that work, the exponent $a\approx
0.5$ was obtained in $1+1$ dimensions).

It is possible to extend the scaling arguments of Sec. III to explain why
$a=1/2$ for BBD in all dimensions. As discussed above, the correlation between
columns is produced by the incidence of large particles at time intervals of
order $1/F$, for small $F$, and the random growth leads to a typical height
difference between neighboring columns of order $1/F^{1/2}$. Thus, when a large
particle is deposited, it will typically create a
pore with height of order $1/F^{1/2}$ and with lateral size of order unity. This
is confirmed in Figs. 6(a) and 6(b), where we show cross sections of deposits with
$F=1$ and $F=0.2$: for the smaller value of $F$ [Fig. 6(b)], the pores are narrow
in the horizontal direction, but vertically high. The
overall balance for small $F$ is that an empty volume of order $1/F^{1/2}$ is
produced during a
time interval in which $1/F$ solid particles are deposited. Consequently, the
porosity is expected to be 
\begin{equation}
P = V_\textrm{\footnotesize pore}/\left( V_\textrm{\footnotesize solid}+V_\textrm{\footnotesize pore}\right)  
\sim \left( 1/F^{1/2}\right) /
\left( 1/F + 1/F^{1/2}\right) \sim F^{1/2} ,
\label{porosity}
\end{equation}
which gives $a=1/2$. Since this argument is
based on random deposition properties and the stochastic rules of BBD, it is
valid in all substrate dimensions.

The above approach is interesting because it shows that the properties of
the surface of the deposit, which depend on the particular growth rules of the
model, can be used to determine scaling properties of the porous media below it.
A more refined approach was recently used to calculate density-density
correlations in the original BD model \cite{strucbelow}, which also illustrate
the possibility of connecting surface and bulk properties.
These results may motivate the related studies in systems with more complex
growth dynamics.

\section{Connectivity and fractal dimension of the pore system}

An interesting property of the deposits generated by BBD is the fact that they
have a highly connected pore structure even for relatively small values of the
flux of large particles, which are responsible for the formation of pores.
This is illustrated in Fig. 6(b), where the deposit with $F=0.2$ still shows a
high connectivity of the internal pore system.

In order to decide whether the internal pore system percolates (i.e., it is
connected from the substrate up to the external surface), we had to generate
relatively small deposits due to memory restrictions. For substrates of linear
size $L=256$, we
typically deposited $150$ monolayers of particles, which may lead to average
heights of the deposits near $500$ units. Despite these limitations, we were
able to observe the transition from an open (percolating) pore structure to a
structure of isolated (closed) pores in a narrow region of the parameter $F$.
Figure 7 shows the probability of
percolation $P_P$ as a function of $F$, which indicates a percolation transition
at $F_c=0.040\pm 0.005$. In the limit of infinitely large system sizes, this
probability is expected to be a step function with discontinuity at $F_c$, since
it is the probability that a given configuration percolates ($P_P$ must not be
confused with the order parameter of a percolation problem, which is defined as
the fraction of the pores belonging to the percolating cluster and which
continuously decrease to zero at $F_c$). 

The percolation transition
in BBD deposits takes place for porosities between  $20\%$ and $25\%$ (see Fig.
6). This is slightly below the critical probability of percolation with
randomly distributed vacancies in a simple cubic lattice, $p_c=0.3116$
\cite{aharonystauffer} -- in that case, this probability is equal to the
porosity. On the other hand, those porosities are still much higher than the
ones obtained at the percolation transitions in grain consolidation models,
which are close to $0.03$ \cite{roberts}. Those models do not account for
particle deposition processes, but lead the formation of narrow channels in a
very compact structure by allowing the expansion of internal grains.  

In Ref. \protect\cite{yu1}, the properties of the internal and the external
surface of porous deposits produced by BD were studied. In the three-dimensional
case [$(2+1)$-dimensional growth], the deposits grown with a model where a
fraction of the incident particles could relax after deposition were also
studied. Compact layers were obtained for $p<0.35$, where $p$
is the probability that the incident particle diffuses to a smaller neighboring
column, while $1-p$ is the probability of aggregation according to the rules of
the original BD model. Low values of porosity (of order $20\%$) were
also obtained near the percolation threshold there. However, the most important
conclusion that can be drawn from comparison of these models is that the
suppression of relaxation mechanisms have a drastic effect in increasing the
pore connectivity if lateral aggregation takes place with low rate, which is
the case of BBD with small $F$.

Now we analyze the properties of the set of surface sites, which are defined
as the sites of the deposit which have at least one nearest neighbor belonging
to the percolating pore system. This set may be viewed as the internal surface
of the pore structure. The total surface area is expected to scale with the
coverage $\theta$ (number of deposited monolayers) as \cite{yu1} 
\begin{equation}
S \sim L^2\theta^{D_F-D} ,
\label{scalingS}
\end{equation}
where $D_F$ is the fractal dimension of the internal surface and $D=2$ is the
substrate dimension.

In Fig. 8 we show $\ln{\left( S\theta^2/L^2\right)}$ versus $\ln{\theta}$ for
BBD with $F=0.1$ and $F=0.6$. In both cases we obtain $D_F\approx 2.9$ (smaller
error bars are near $3\%$ for $F\sim 0.1$). This fractal dimension is the same
obtained in deposits generated with the original BD model and with the model
including relaxation in Ref. \protect\cite{yu1}. It strongly suggests
universality of $D_F$ among
ballistic deposition models in two-dimensional deposits. Again, these
numerical findings may be viewed as motivation for further theoretical
investigation of the relations between surface growth and bulk properties.

\section{Summary and conclusion}

We studied the surface roughness scaling in the bidisperse ballistic deposition
(BBD) model in $2+1$ dimensions and the properties of the pore system in the
deposits generated with that model.

Systematic extrapolation of roughness and dynamical exponents obtained from
simulation in a range of system sizes showed that it presents KPZ scaling for
fluxes of large particles $F\geq 0.1$. The values of $F$ analyzed here include
the region where previous work suggested a transition in roughness scaling.
Comparison of roughness distributions provided additional support to the
conclusion that the model is in the KPZ class for all values of $F$. A scaling
approach was used to predict the characteristic time of the crossover from
random to correlated growth in BBD and provided relations between scaling
amplitudes and $F$ in the KPZ scaling regime. The theoretically predicted
crossover exponents were also confirmed by simulation. This scenario rules out
the possibility of a roughening transition in BBD. 

The porosities $P$ of the deposits generated with BBD were measured for various
values of $F$ and it was shown that, for small $F$, it increases as $P\sim
F^{1/2}$. This result is explained with an extension of the previous scaling
approach. The deposits have high connectivities in a large range of values of
$F$, down to $F\approx 0.05$. The comparison with ballisticlike models
involving surface relaxation show that the suppression of relaxation
significantly enhances the connectivity of the pore system for low rates of
lateral aggregation. The fractal dimension of the internal surface of the
percolating deposits is $D_F\approx 2.9$, which is very close to previously
studied ballisticlike models, thus suggesting universality among these
systems. 

Concerning roughness scaling, the comparison of former results on BBD
\cite{bbd2} and our
work illustrates the typical difficulties involved in finding dynamical
transitions in growth models and the relevance of theoretical approaches which
can provide a deeper insight into the system behavior, even if only in a
qualitative way. On the other hand, this study confirms that systematic methods
to extrapolate simulation data from finite systems can be very helpful tools. 

Concerning bulk features, we believe that our results strongly motivates further
theoretical investigation on relations between surface growth rules and
internal properties of the deposits. Since simple arguments are capable
of explaining the scaling of quantities such as the porosity
in a restricted range of the model parameters, it is expected that more
sophisticated approaches will be able to address the same question in more
complex systems, such as done in Ref. \protect\cite{strucbelow}. Moreover, the
apparently universal surface fractal dimension of ballisticlike deposits is
another important point for further investigation and suggests possible
applications of this class of growth model. Indeed, the estimate $D_F= 2.9$ is
very close to the value $D_F\approx 2.85$
obtained experimentally in sandstones \cite{krohn,radlinski}, and not very far
from $D_F=2.7-2.8$ obtained in gold vapor deposition \cite{gomez}. Another
interesting question is what happens in ballisticlike models with varying
angles of deposition, which have been recently applied to growth of silicon or
silicon compounds \cite{levine,yanguas,fanlike}.

Our conclusions also confirm the relevance of BBD as a simplified but
realistic model for sedimentary rock formation, as originally proposed in Ref.
\protect\cite{bbd1}. In order to provide a quantitative description of real
systems, this type of model may eventually include mechanisms such as those of
the grain consolidation models \cite{roberts}, which allow the expansion of
internal grains, as well as the aggregation of different grain sizes.

\begin{center}
\bf ACKNOWLEDGEMENTS
\end{center}

F.A.S. acknowledges support from CNPq and F.D.A.A.R. acknowledges support from CNPq and
FAPERJ (Brazilian agencies).

\newpage


\newpage

\begin{figure}[!h]
\includegraphics[width=.6\textwidth]{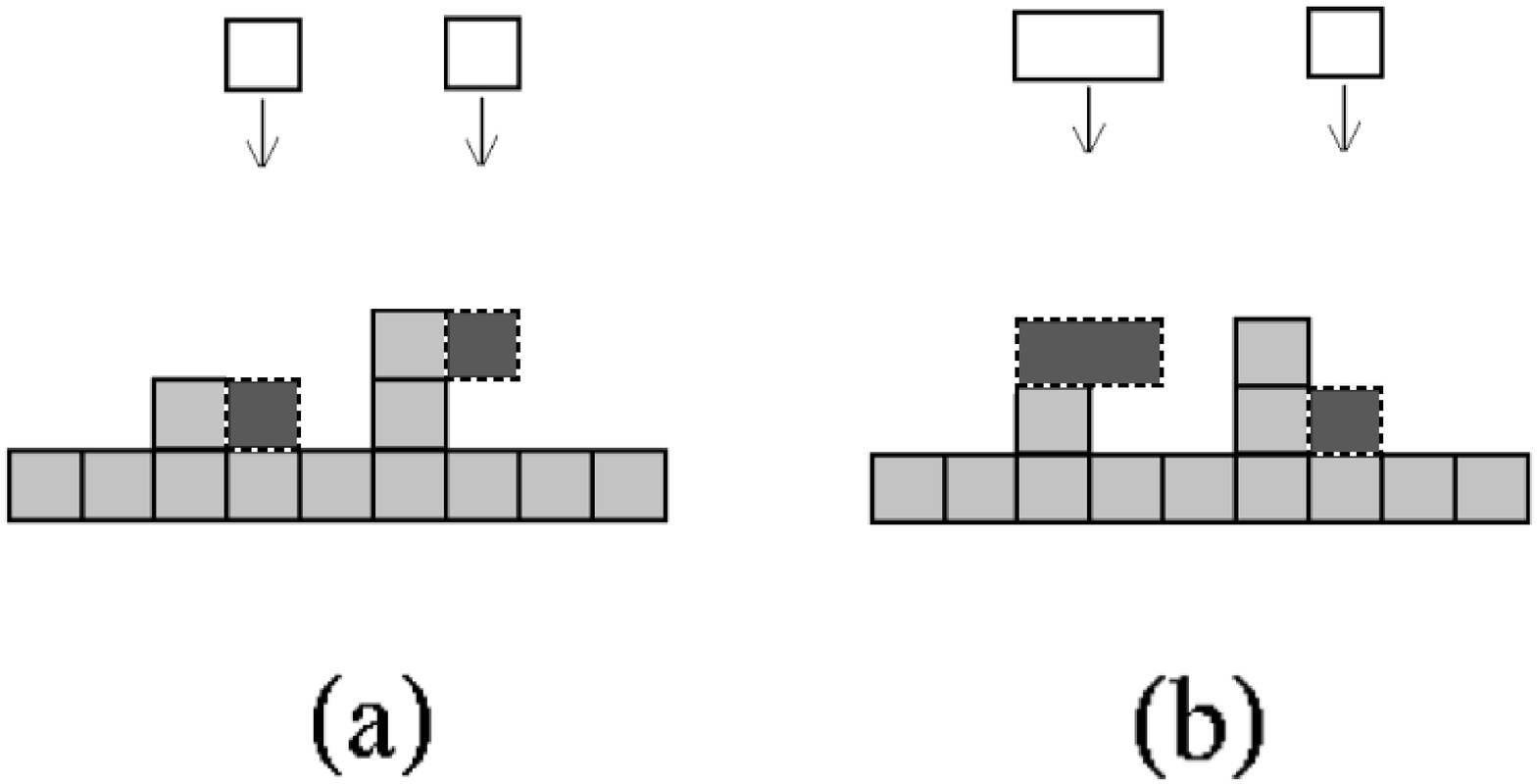}
\caption{(a) Growth rules of the original ballistic deposition (BD) model in a
line. (b) Growth rules of the bidisperse ballistic deposition (BBD) model in a
line. In both cases, aggregated particles are shown in light grey, incident
particles in white and aggregation positions of incoming particles in dark
grey and with dashed contour.}
\label{fig1}
\end{figure}

\begin{figure}[!h]
\includegraphics[width=.8\textwidth]{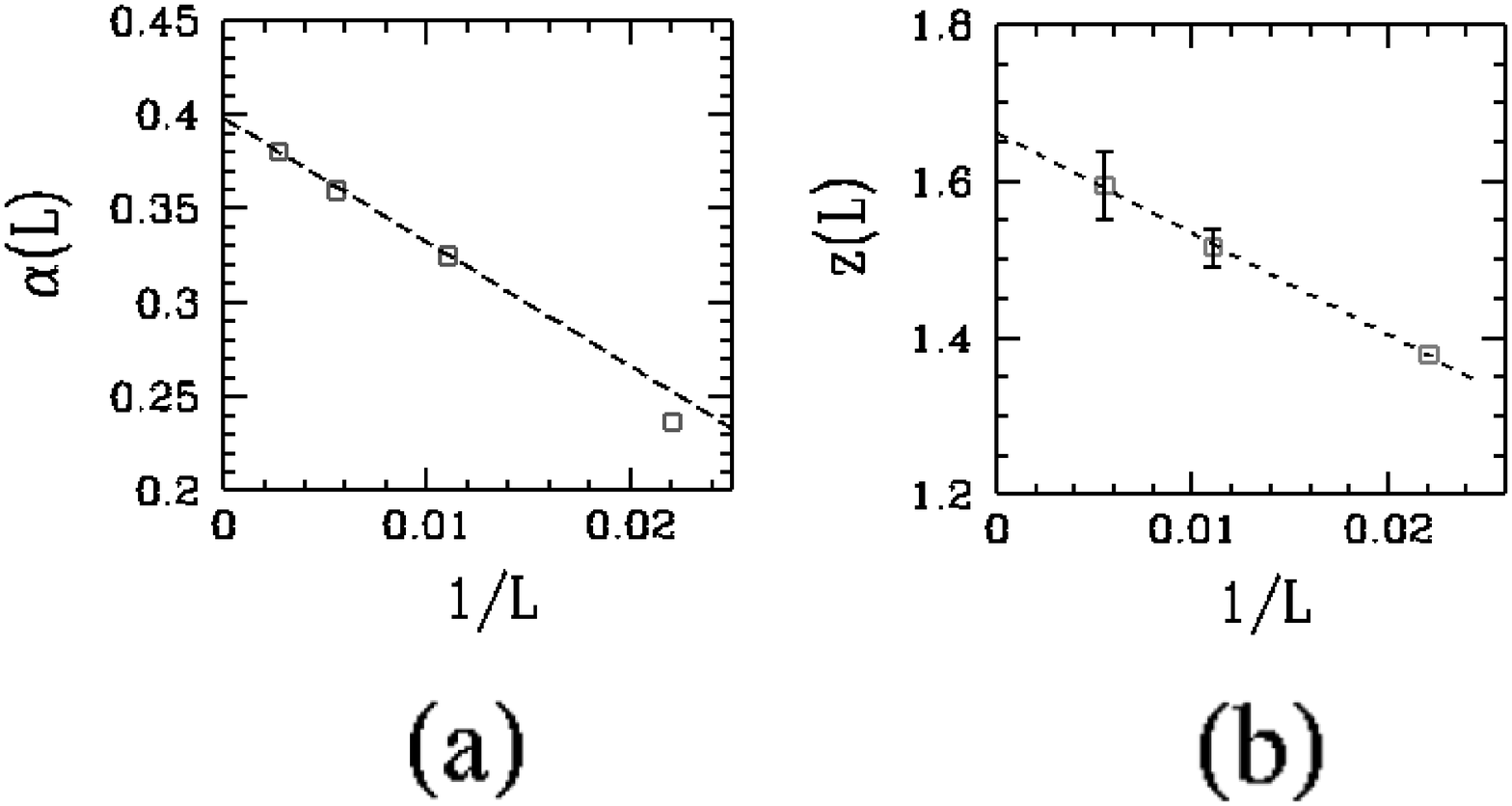}
\caption{(a) Effective roughness exponents versus inverse lattice size for BBD
with $F=0.08$. The linear fit of the data for $64\leq L\leq 512$ (dashed line)
gives $\alpha\sim 0.395$. (b) Effective dynamic exponents versus inverse
lattice size for BBD with $F=0.1$. The linear fit of the data for the largest
lattice sizes (dashed line) gives $z\sim 1.66$. Where error bars are not shown,
they are smaller than the data points.}

\label{fig2}
\end{figure}

\begin{figure}[!h]
\includegraphics[width=.4\textwidth]{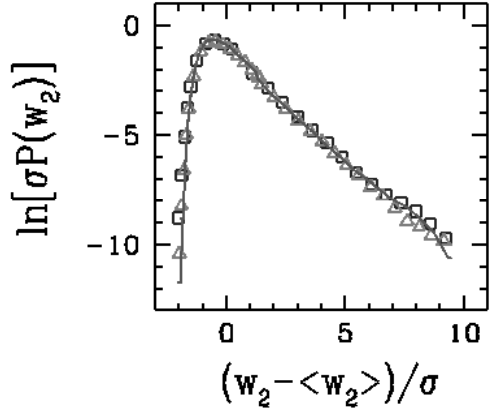}
\caption{Steady state scaled roughness distributions for the BBD model with
$F=0.06$ (triangles) and $F=0.10$ (squares) and for the RSOS model (solid
curve).}

\label{fig3}
\end{figure}

\begin{figure}[!h]
\includegraphics[width=.8\textwidth]{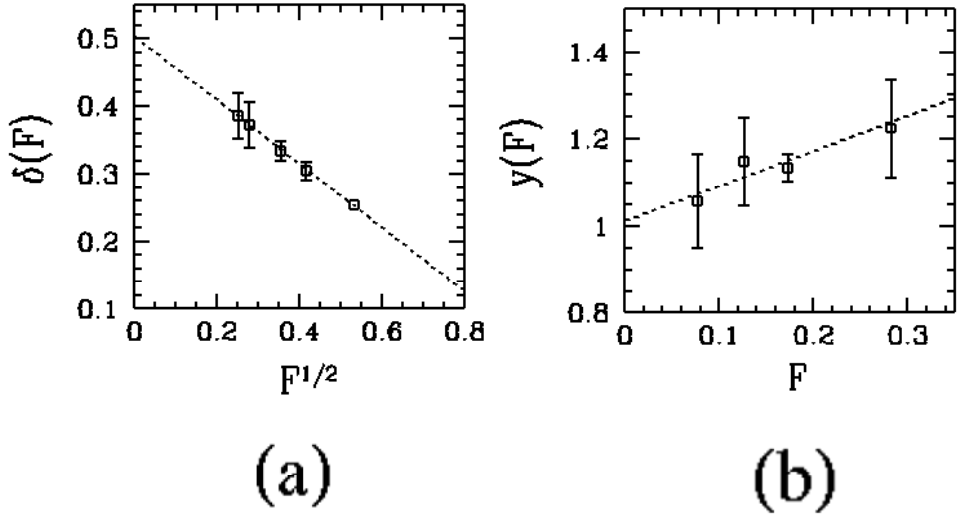}
\caption{(a) Effective exponent $\delta(F)$ as a function of $F^{1/2}$ and a
linear
fit (dashed line) which gives $\delta=0.50\pm0.02$ as $F\to 0$. (b) Effective
exponent $y(F)$ as a function of $F$ and a linear fit (dashed line) which gives
$y=1.01\pm0.10$ as $F\to 0$.}

\label{fig4}
\end{figure}

\begin{figure}[!h]
\includegraphics[width=.4\textwidth]{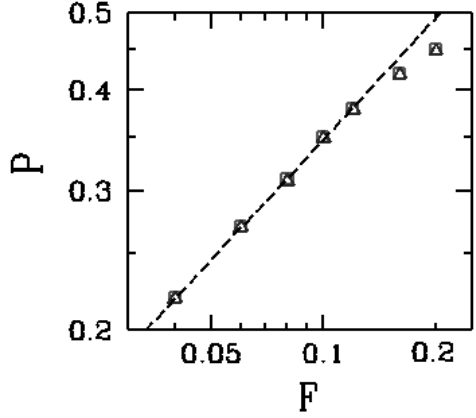}
\caption{Porosity $P$ versus $F$ for BBD in substrate
sizes $L=256$ (squares) and $L=128$ (triangles). The linear fit of the data for
$F\leq 0.1$ (dashed line) has slope $0.503\pm 0.003$.}
\label{fig6}
\end{figure}

\begin{figure}[!h]
\includegraphics[width=.8\textwidth]{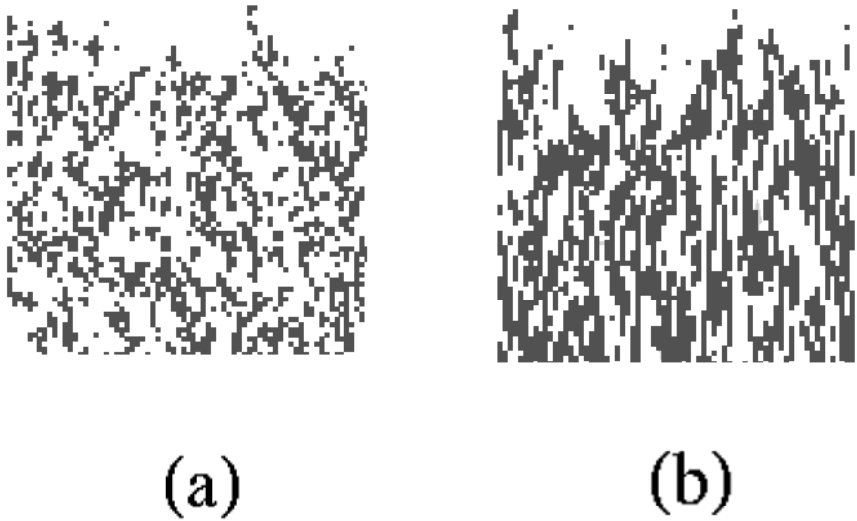}
\caption{Cross sectional view of three-dimensional BBD deposits generated with
(a) $F=1$ and (b) $F=0.20$.}
\label{fig5}
\end{figure}

\begin{figure}[!h]
\includegraphics[width=.4\textwidth]{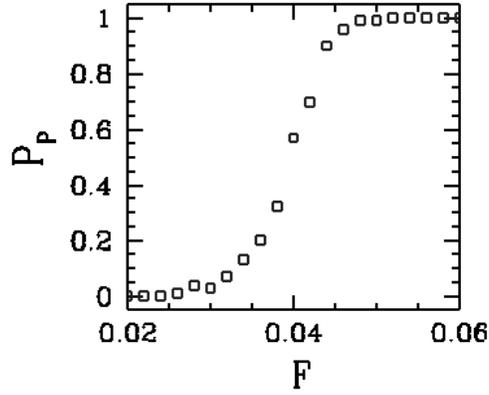}
\caption{Percolation probability $P_P$ versus $F$ for the BBD model in substrate
sizes $L=256$, which indicates  a percolation transition at $F_c\approx 0.04$.}
\label{fig7}
\end{figure}

\begin{figure}[!h]
\includegraphics[width=.4\textwidth]{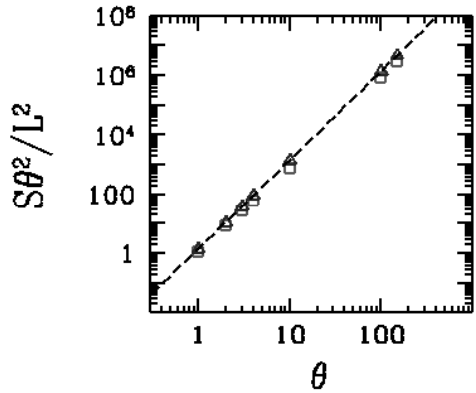}
\caption{Scaled surface fractal dimension as a function of coverage in the BBD
model with $F=0.1$ (squares) and $F=0.6$ (triangles). The linear fits of the
data in the scaling
regions (dashed line) give $D_F\approx 2.9$ in both cases.}
\label{fig8}
\end{figure}

\vfill

\end{document}